\documentclass[11pt, twoside, letterpaper]{article}
\usepackage[margin=1.25in]{geometry}
\usepackage{booktabs}
\usepackage{makecell}

\usepackage{fancyhdr}
\usepackage{amsmath}

\usepackage[utf8]{inputenc} 

\usepackage[sort,compress]{natbib}
\setcitestyle{numbers}

\usepackage[format=plain,labelfont={bf},textfont={small,it}]{caption}
\usepackage{graphicx}

\usepackage{subcaption}
\DeclareCaptionLabelFormat{ms_label_style}{\sffamily\itshape\small\linespread{1}\selectfont#2}
\DeclareCaptionFont{ms_font_style}{\sffamily\itshape\small\linespread{1}\selectfont}

\usepackage{totcount}
\newtotcounter{citnum} 
\def\oldbibitem{} \let\oldbibitem=\bibitem
\def\bibitem{\stepcounter{citnum}\oldbibitem}

\graphicspath{{figs/}}

\title{\LARGE\bfseries Algorithmic Bias in Lending: Evidence from a Fintech Audit}

\author{
  Madison Coots\\
  Harvard University
  \and
  Robert Bartlett\\
  Stanford University
  \and
  Julian Nyarko\\
  Stanford University
  \and
  Sharad Goel\\
  Harvard University
}

\date{}

\begin{document}
\maketitle

\thispagestyle{empty}

\begin{abstract}
Algorithmic lending has transformed the consumer credit landscape, with machine learning models commonly facilitating underwriting decisions. 
To comply with fair lending laws, these algorithms exclude legally protected characteristics, such as race and gender.
Yet algorithmic underwriting can still inadvertently favor certain groups, 
prompting concerns about whether lending algorithms exhibit discriminatory behavior.
Using proprietary loan-level data from a major U.S. fintech platform, we audit lending decisions across approximately 80,000 personal loans.
We find that loans made to men and Black borrowers yielded lower profits than loans to other groups, suggesting that men and Black borrowers benefited from relatively favorable pricing. We trace these disparities to miscalibration in the platform's underwriting model, which overestimates risk for women and underestimates risk for Black borrowers. 
We then show that one could correct this miscalibration---and the corresponding disparities---by including race and gender in underwriting models, illustrating a tension between competing notions of fairness.
\end{abstract}

\section{Introduction}
\label{intro}
Fifty years ago, when the Fair Housing Act (FHA)~\cite{USCongress1968} and the Equal Credit Opportunity Act (ECOA)~\cite{USCongress1974} established the foundation for fair lending regulation, the lending landscape looked very different from today. At that time, borrowers typically applied for loans in person at a bank, where approval and pricing decisions were made at the discretion of individual loan officers. This discretion was widely viewed as a potential source of discrimination, raising concerns about whether implicit or explicit biases against racial minorities and women influenced lending decisions. 

Over the last twenty years, however, the lending landscape has been transformed by the rise of algorithmic lending---and  ``fintech lending'' in particular. 
Under the fintech model of lending, borrowers apply for loans online, and underwriting is handled by sophisticated machine learning models rather than individual lending officers, reducing or even eliminating human discretion.
To comply with fair lending regulation, these models exclude protected characteristics and their proxies~\cite{campbell2019making}.
The rapid growth of algorithmic lending has largely been motivated by the financial benefits of more accurately predicting creditworthiness, by using highly complex underwriting models in conjunction with non-traditional predictors of risk~\cite{DiMaggio2022, DiMaggio2020}. Similarly motivated traditional lenders have likewise  adopted complex, machine learning-based approaches to underwriting, significantly increasing the prevalence of algorithmic lending across the industry.

The growing shift from human to algorithmic decision making has increased concerns about the risks of \textit{algorithmic} bias in the models used to make lending decisions~\cite{barocas2016big, avery2012does, hurley2016credit, hurlin2024fairness, fuster2017predictably}. These concerns largely stem from previous findings of algorithmic biases against racial minorities and women in other consequential domains, including in healthcare \cite{obermeyer2019dissecting}, criminal risk assessments \cite{skeem2016gender}, and hiring \cite{lambrecht2019algorithmic, gaebler2024auditing}. Concerns about algorithmic bias in underwriting models have in turn prompted scholars to revisit questions about how best to detect and regulate discrimination in the fintech era~\cite{Kumar2022, ChohlasWood2023, Kleinberg2018, Barocas2019, Gillis2022}, building upon a large body of empirical research on detecting lending discrimination more generally~\cite{Giacoletti2023, Park2021, Munnell1996, Bhutta2024, Bhutta2020, Dobbie2021, ravina2019love}. 

Despite these growing concerns, algorithmic bias in fintech lending has proven challenging to measure empirically. These challenges arise principally from the fact that researchers and regulators rarely have had access to the full set of relevant (often proprietary) information used by fintech lenders to make approval or pricing decisions. Consequently, the most common approach for assessing the fairness of algorithmic lending decisions has been to consider ``adverse impact ratios'' (AIRs): the ratio of the decision rate (e.g., loan approval or interest rate) for members of a protected group relative to a reference group, such as men or non-Hispanic White borrowers.
However, relatively low approval rates or high interest rates may simply reflect differences in creditworthiness between groups, rather than discrimination, limiting the probative value of AIRs. Recognizing this limitation of AIRs, researchers have also sought to audit fintech lenders by estimating differences in approval rates and interest rates across groups after adjusting for observable risk factors in a regression model~\cite{Bartlett2022, Bhutta2020, Bhutta2024}. 
This approach, however, likewise suffers from significant limitations, both practical and conceptual. Failing to adjust for the complete set of information used to make lending decisions can produce misleading results and may over- or understate discrimination~\cite{ayres2002outcome}. This potential for omitted-variable bias is even greater with fintech lending, where underwriting models often rely on thousands of proprietary covariates.
Further, this regression-based approach can at best detect the \textit{direct} effects of protected attributes on decisions, limiting its utility for auditing facially neutral lending models that do not explicitly condition decisions on race or gender~\cite{Jung2024}. In this case, audits that adjust for the full set of risk variables would (correctly) find no marginal effect of race or gender on decisions. Yet even with facially neutral algorithms, lenders may inadvertently discriminate against certain groups. For example, systematic errors in estimating risk may lead lenders to favor or disfavor members of certain groups---an algorithmic variant of redlining~\cite{CorbettDavies2023}.

In this work, we conduct a large-scale audit of algorithmic lending decisions using novel, proprietary loan-level data drawn from a large online fintech platform in the U.S. In contrast to prior work, our audit is based on data that includes the full set of proprietary covariates used in the lender's underwriting model, providing visibility into the mechanics of the lender's pricing decisions. We are primarily concerned with measuring discrimination in loan pricing, as defined under U.S. fair lending law---which prohibits disparities in loan pricing if they do not arise from differences in credit risk. As a matter of U.S. federal law, fair-lending obligations prohibiting discriminatory practices arise principally under the Fair Housing Act (FHA) and the Equal Credit Opportunity Act (ECOA). U.S. courts have determined these statutes prohibit two distinct types of discriminatory conduct~\cite{taylor2018ecoa}. For one, they render illegal ``disparate treatment,'' which encompasses lending decisions that are intentionally based on a protected characteristic, such as race or gender. Additionally, these laws go further and also prohibit certain facially neutral lending practices that create disparities in ways that correlate with protected characteristics. This latter form of discrimination is referred to as ``disparate impact,'' and it is the focus of our analysis. To state a disparate-impact claim, a plaintiff must first identify a specific policy or practice that produces a statistically significant disparity between protected and unprotected groups. This showing shifts the burden to the defendant-lender to demonstrate that the challenged practice is necessary to achieve a legitimate and nondiscriminatory objective. Importantly, the only type of business objective that courts have recognized as sufficient to justify disparities in lending outcomes is creditworthiness.\footnote{See A.B. \& S. Auto Service, Inc. v. South Shore Bank of Chicago, 962 F.Supp. 1056 (N.D. Ill. 1997) (“[In a disparate impact claim under the ECOA], once the plaintiff has made the prima facie case, the defendant-lender must demonstrate that any policy, procedure, or practice has a manifest relationship to the creditworthiness of the applicant. . . ”); Lewis v. ACB Business Services, Inc., 135 F.3d 389, 406 (6th Cir. 1998) (“The [ECOA] was only intended to prohibit credit determinations based on ‘characteristics unrelated to creditworthiness.”).} Justifications stemming from ``market forces''---such as differences in the existence of credit deserts or other circumstances affecting a borrower's willingness to pay---have explicitly been deemed illegitimate reasons incapable of justifying a difference in loan pricing.\footnote{See Miller v. Countrywide Bank, NA, 571 F.Supp.2d 251, 258 (D. Mass 2008) (rejecting argument that discrimination in loan terms among African American and White borrowers was justified as the result of competitive “market forces,” noting that prior courts had rejected the “market forces” argument insofar that it would allow the pricing of consumer loans to be “based on subjective criteria beyond creditworthiness."} 
Though these issues have not yet been litigated in the context of algorithmic pricing specifically, we expect these legal principles extend to such pricing systems.
Therefore, consistent with this legal framework, we define discriminatory loan pricing as pricing disparities that correlate with protected characteristics and that cannot be explained by differences in creditworthiness.

In our audit, we employ a profit-based test of lending discrimination, first articulated by Becker~\cite{Becker1971, Becker1993}. This test relies on straightforward intuition: If a lender prices loans accurately, then---after setting interest rates to reflect risk---we would expect loans to be similarly profitable across groups. Indeed, in a competitive marketplace comprised of risk-neutral lenders, we would expect loans to be priced to achieve the same expected return for every individual borrower.
Consequently, if loans made to a particular group are systematically more profitable, this suggests that the lender priced those borrowers too aggressively relative to their true risk, consistent with discrimination. To compare profitability, we rely on the annualized internal rate of return (IRR), a common measure of profit in lending that directly connects loan pricing with repayment outcomes. In practice, lenders are not perfectly risk neutral, and may expect a premium to lend to riskier borrowers. We explore this point further in our analysis. 

In estimating the annualized IRR across racial and ethnic and gender groups, we find that loans made to Black borrowers are less profitable than those to other racial subgroups; and that loans made to men are less profitable than those made to women. These results suggest the lender's decision algorithm systematically  benefits Black borrowers and men. We confirm this apparent benefit by tracing these disparities in part back to miscalibration in the lender's underwriting model, which underestimates the borrowing risk of Black borrowers and overestimates the risk of women. One can correct this miscalibration by explicitly including race and gender in risk models, but doing so would violate fair lending laws, illustrating a tension between competing notions of fairness.

\section{Lending Data}
Our study uses data from a fintech platform that facilitates lending to individuals for a variety of purposes. The platform employs a proprietary underwriting model that utilizes both traditional and non-traditional information to
assess lending risk and originate unsecured personal loans. Both the loan approval and pricing decisions are fully automated, with no human intervention.

The platform we study provided two distinct, non-overlapping sets of data. The first dataset consists of 79,251 funded loans from 2019 (i.e., loans that were approved and originated). The second dataset consists of 125,345 3-year loan applications from 2021 to 2023 (i.e., a mix of both approved and denied loan applications, as well as loans approved by the lender, but not accepted by the applicant). For each individual in the data, we observe the application date, as well as the approval status and terms (amount and APR) of each loan for which the applicant was considered. We additionally observe the full set of proprietary risk variables used by the lender to estimate default risk at the time of application. For the set of funded loans only, we observe the target returns (i.e., profit targets) set by the lender for each loan. Our primary analysis is based on 79,251 3-year loans that were approved and originated in 2019.
For these funded loans, we additionally observe both the principal and the borrower's monthly loan repayments.  Summary statistics, including the demographic composition of each data set, are presented in Tables \ref{tab:funded_borrowers} and \ref{tab:applicants}.

Lenders are legally prohibited from collecting information on applicants' protected characteristics and so, in line with regulatory practices~\cite{cfpb2014_bisg}, applicant's demographics are inferred from their names to facilitate auditing. 
Specifically, the platform used the Bayesian Improved Surname Geocoding (BISG) proxy methodology \cite{Elliott2008, Elliott2009} to estimate the probability that each individual was White, Black, Hispanic, Asian, or ``other.'' Gender estimates were similarly derived from applicants' first names.

For all analyses that use BISG and gender probabilities as weights, we conduct robustness checks that reproduce our results using the argmax of the probability estimates to obtain a single, most likely race and gender label for each individual (see Appendix). That approach is in line with recent recommendations in the fair lending literature \cite{zhang2018assessing}.
The results from this alternative analysis qualitatively mirror those from our primary approach.

\section{Results}
\subsection{Estimates of profit by race and gender}

Using the set of funded loans, we start by computing the realized profit of the aggregate loan portfolio of each race and gender subgroup.
Specifically, for each subgroup, we first combine the individual loan repayment histories into an aggregate cash flow vector that indicates the net amount the lender received from borrowers each month. Vector entries may be either positive or negative---with negative entries indicating net loan dispersal that period.
To measure profit, we compute the annualized internal rate of return (IRR) of each aggregate cash flow, a metric commonly used by lenders to characterize loan profitability. The annualized IRR is the annualized rate that makes the net present value of all loan repayments equal to the loan principal disbursed in the first period. IRR is scale-invariant (i.e., its magnitude does not depend on the size of the loan or portfolio of loans) and it incorporates the time value of money. IRR can range from –100\% (total loss) through 0\% (break-even) to arbitrarily large positive values (the faster or larger the repayments relative to the principal, the higher the IRR). 

\begin{figure}
\centering
\includegraphics[width=0.6\linewidth]{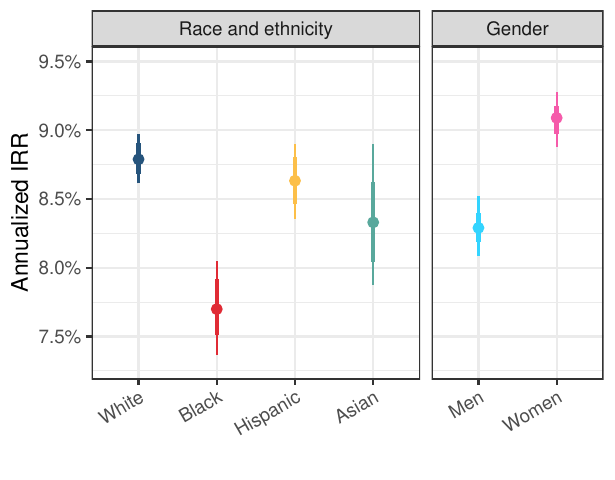}
\caption{\textbf{Results of computing the annualized IRR of cashflows aggregated across race and gender groups.} Relative to other groups, the lender earns less profit on loans made to Black borrowers and men, suggesting these groups benefit from relatively favorable loan terms. We show the 68\% confidence interval (thick bar) and the 95\% confidence interval (thin bar) of the IRR estimates.}
\label{fig:irr}
\end{figure}

In Figure \ref{fig:irr}, we plot estimates of profit by subgroup.\footnote{Recent work has suggested that the intersection of race and gender is an important dimension to consider when evaluating lending disparities across groups \cite{harkness2016discrimination}. In Figure~\ref{si:intersection}, we show estimates of annualized IRR for intersectional race–gender groups (e.g., Black women, Black men). The results are qualitatively similar to those shown in Figure \ref{fig:irr}, revealing that loans made to men within a given race group earn less profit than those made to women.}
We find that loans made to White, Hispanic, and Asian borrowers earn similar returns (between 8.3\% and 8.8\%), whereas loans made to Black borrowers earn noticeably less (7.7\%). Likewise, the profits of loans made to men (8.3\%) fall below those of women (9.1\%). These results thus suggest the platform's algorithmic pipeline ultimately benefits Black borrowers and men, offering them relatively favorable loan terms.

We also estimated group-IRRs a second way, instead by computing the IRR of individual loan cashflows and then taking a weighted average of individual IRRs using the race and gender probabilities as weights.
Individual cash flows may have a mathematically undefined IRR if the cash flow reflects an immediate default (i.e., no payments are observed after the loan is disbursed) or an immediate prepayment (i.e., the loan is prepaid in the same period it is disbursed). In the case of immediate defaults, we assumed an IRR of -100\% to reflect a total loss. To address issues stemming from immediate prepayments, we constructed all loan cash flows so that the first observed payment occurs in the period immediately following loan disbursement. This way, a loan cannot be prepaid in the same period in which it was disbursed, and the resulting cash flow yields a well-defined IRR.
Both approaches of estimating group IRRs yielded qualitatively similar results. The results of this approach for computing group profits are
discussed further in Section~\ref{section:shopping}, but are qualitatively similar to those shown in Figure~\ref{fig:irr}.

\paragraph{Default rates and the problem of inframarginality.} Rather than examining differences in profit to identify discriminatory lending, one might instead consider differences in default rates. In particular, mirroring the logic of a profit-based test, one might reason that a group of borrowers with relatively high observed default rates benefited from a less stringent lending standard,  being granted loans even though they were relatively high risk.
That approach, however, suffers from the problem of inframarginality~\cite{ayres2002outcome,simoiu2017problem,gaebler2025}. To see this, suppose that a lender perfectly estimates default risk for all applicants based on the available application data, and that loans are subsequently granted to all applicants with risk below a certain threshold, say 10\%. In that scenario, lenders would be behaving efficiently---and would not be violating fair lending laws---yet, in general, we would expect group-specific default rates to differ. Indeed, the group-specific default rate is the conditional average of default risk for all members below the lending threshold. If risk distributions differ across groups, we would accordingly expect default rates among loan recipients to likewise differ, even in the absence of discrimination.
A profit-based outcome test, however, mitigates this concern. Assuming risk neutrality and a competitive marketplace, we would expect lenders to price loans to achieve identical expected returns from every applicant. If a lender's risk estimates are accurate, the average realized returns across subgroups would accordingly be similar, even if risk profiles differ across groups, sidestepping the problem of inframarginality~\cite{ayres2002outcome}.

\subsection{Accounting for Risk Aversion}
Interpreting differences in group-level profits as evidence of discrimination assumes lenders are risk neutral.
Potential risk aversion complicates the logic, since we expect a risk averse lender to demand a premium from higher risk borrowers. As a result, differences in risk across groups could lead to differences in IRR, even in the absence of discrimination---a manifestation of the problem of inframarginality.

\begin{figure}
\centering
\includegraphics[width=.5\linewidth]{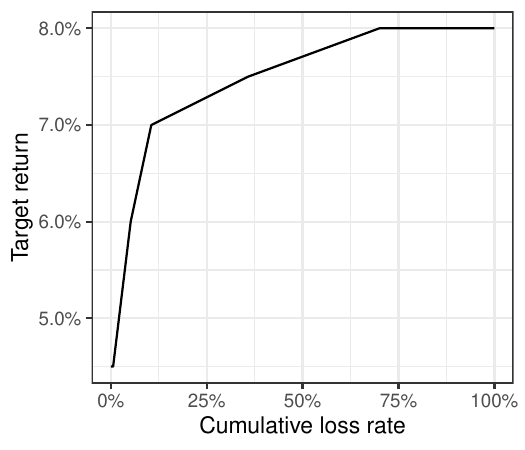}
\caption{\textbf{An example target return curve used by the lender to price loans.} The cumulative loss rate is the 
proportion of the principal that the lender expects to lose.
The curve indicates a preference for relatively higher returns for riskier loans, consistent with risk aversion.}  
\label{fig:target-return}
\end{figure}

Indeed,
the lender we consider explicitly prices loans to achieve a target return that increases as a function of risk.
Figure~\ref{fig:target-return} displays a sample target return curve provided to us by the lender, which operationalizes risk in terms of the ``cumulative loss rate'': the proportion of the principal that is expected to go uncollected from a prospective borrower. Formally, the cumulative loss rate is:
\begin{equation}
\frac{1}{P_0} \sum_t f(t) \cdot (P_t - R_t),
\end{equation}
where $P_0$ is the original principal; for each period $t$, $f(t)$ is the probability of default in that period; $P_t$ is the remaining principal; and $R_t$ is the expected recovery conditional on default. The cumulative loss rate thus captures the likelihood and timing of defaults, as well as expected recovery amounts in the case of such events.

\begin{figure}
\centering
\includegraphics[width=0.65\linewidth]{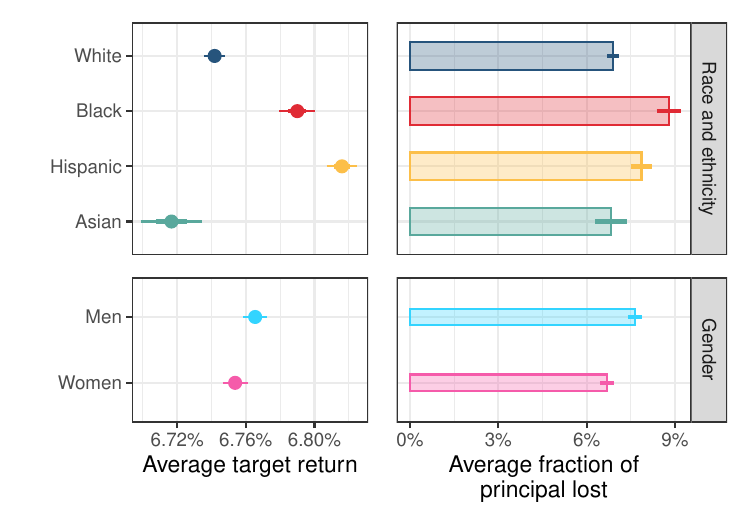}
\caption{\textbf{Average target returns and the average fraction of principal lost across race and gender groups.} (Left) The lender has higher target returns for Black, Hispanic, and men borrowers than other groups. Target returns are set by the lender as a function of the estimated cumulative loss rate for each loan, as shown in Figure \ref{fig:target-return}. Differences in group average target returns therefore correspond to differences in the average estimated cumulative loss rates across groups. We show the 68\% confidence interval (thick bar) and the 95\% confidence interval (thin bar) of the estimates of average target returns within each group. (Right) The lender loses greater fractions of loan principals to Black borrowers and men. The thick line shows the 95\% confidence interval of our estimates.}
\label{fig:target_return_principal_lost}
\end{figure}

In our analysis above, we observed that Black and male borrowers were less profitable. This pattern would be consistent with non-discrimination if it were also the case that these groups were lower risk than their non-Black and female counterparts---and thus the lender intentionally sought lower returns from them. We empirically investigate this possibility in two ways.
First, we consider the lender's stated target return for each borrower. We display the average target returns across groups in the left panel of Figure~\ref{fig:target_return_principal_lost}, and find that Black and male borrowers are among the groups for which average target returns are highest---not lowest. 
Second, in the right panel of Figure~\ref{fig:target_return_principal_lost}, we compute the observed proportion of principal lost by group, similarly finding that Black and male borrowers are among the highest risk groups.
(We assume for simplicity that $R_t = 0$, meaning that no additional principal was recovered after default.)
It therefore appears that risk aversion does not explain the profit gaps we observe. If anything, we would expect, in the absence of discrimination, that Black and male borrowers would be slightly more, not less, profitable.\footnote{%
In practice, target returns may not be readily available to auditors to conduct the type of analysis we carry out above; and, even if they were available, auditors may not wish to rely on such lender attestations.
While cumulative loss can be approximated by loan repayment data, auditors might instead consider using 
observed default rates as a simple, more easily interpreted measure of risk. 
Figure~\ref{fig:default-rate} shows that Black borrowers and men have higher default rates than other groups, in line with the cumulative loss comparisons. Moreover, figure \ref{si:default-target-returns} shows that default rates and target returns are empirically strongly correlated, further illustrating the value of considering default rates when accounting for risk aversion. 
Auditors might also examine differences in the volatility of IRR across groups as an additional indicator of group-level risk, as shown in figure~\ref{si:volatility}.}

\subsection{Miscalibrated Risk Estimates}

What then drives the differences in profitability across borrowers? 
One possibility is that the lender systematically underestimates the risk of Black and male borrowers, leading to more favorable interest rates relative to their true riskiness. To investigate this possibility, we would ideally assess the calibration of the lender's internal risk score, i.e., comparing predicted and realized default rates. 
However, this proprietary risk score was withheld by the lender. Instead, we evaluate the calibration of our own (race- and gender-blind) risk model trained with the same set of proprietary features used by the lender's internal model.

Our models are trained using XGBoost with 5-fold cross-validation to obtain out-of-sample predictions. Our first model---the ``blind" model---serves as a proxy for the lender's internal underwriting model because it is race- and gender-blind, meaning that it does not use information on race and ethnicity or gender to make risk predictions. Our second model, however---the ``aware" model---is race- and gender-aware and uses the BISG probabilities to estimate risk. Our blind and aware risk models obtained an out-of-sample area under the curve (AUC) of 0.665 (95\% CI, 0.662, 0.669) and 0.667 (95\% CI, 0.662, 0.671), respectively.

\begin{figure}[t]
\centering
\includegraphics[width=0.6\linewidth]{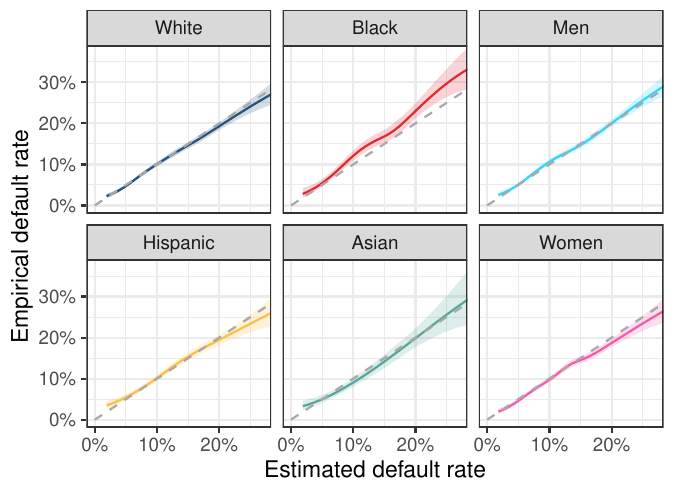}
\caption{\textbf{The calibration of reconstructed race- and gender-blind risk scores.} The line $y=x$ denotes the line of perfect calibration, shown by a dashed gray line. (Left and Middle) When disaggregated across racial and ethnic groups, blind risk scores are miscalibrated for Black borrowers and tend to underestimate their likelihood of default. (Right) Blind risk scores are calibrated for men, but tend to slightly overestimate the risk of women.
These results collectively suggest that inaccuracies in the lender's (blind) risk score lead to relative underpricing of loans for Black borrowers, and relative overpricing of loans for women.}
\label{fig:blind-calib}
\end{figure}

A calibrated underwriting model should have estimated default rates that align with observed default rates. As shown in Figure \ref{fig:blind-calib}, our blind model is well-calibrated for White, Asian, and Hispanic borrowers, but is noticeably miscalibrated for Black borrowers, significantly \textit{underestimating} Black borrowers' risk of default. Disaggregated by gender, we find that the model is in fact well-calibrated for men, but slightly miscalibrated for women, \textit{overestimating} women's default risk.

By approximately reconstructing the lender's risk model, 
our findings suggest that miscalibration at least in part explains the observed differences in profitability across groups.
But our analysis is inherently limited since we do not have access to the actual, proprietary risk model the lender employs.
To address this limitation, we note that we would expect APR to increase approximately monotonically in the lender's internally estimated risk score.
Leveraging this observation, we examine group-specific default rates as a function of APR in Figure \ref{si:interest-rate-fig}.
In the absence of any miscalibration, we would expect to see similar default rates across groups given a fixed APR.
However, we instead find that, for any given APR, Black borrowers default at higher rates than White borrowers, and women borrowers default at lower rates than men. 
This result suggests that the miscalibration we observe in our own blind risk model likewise occurs in the lender's own internal model.
This general strategy of assessing model miscalibration is one that external regulators could also follow in similar audits of fintech lenders, since loan-level APR and default are typically readily available.

While assessing model calibration is relatively straightforward, diagnosing the precise cause is considerably more challenging. 
In general, miscalibration is not unexpected in race- and gender-blind models, since it is possible that groups exhibit differences in risk that remain even after adjusting for the factors included in the model~\cite{CorbettDavies2023}. 
Although miscalibration in blind risk models is not inevitable, its presence does not necessarily imply negligence in model development.
One possible explanation is that the set of legally permissible underwriting variables does not fully capture the factors that are relevant to default risk.
For example, even after controlling for income and other observable financial characteristics, borrowers of different groups may differ in their access to social capital and informal financial support networks that affect their ability to avoid  defaults during periods of financial precarity \cite{burchardi2013economic}. However, this is but one possible explanation of many, making it difficult to pinpoint the exact source of the miscalibration we observe in our analysis.
Nevertheless, the direction of the miscalibration that we observe in the lender's risk model is consistent with other work that has observed similar patterns with other blind estimates of risk. For example, credit scores have been found to underestimate risk for Black borrowers~\cite{bakker2025credit}, and  criminal risk assessments tend to overestimate recidivism risk for women~\cite{skeem2016gender}. 

One way to statistically correct the subgroup miscalibration is to explicitly include race and gender in risk models. In Figure \ref{si:aware-calib}, we plot the calibration of our race- and gender-aware model, and confirm that risk estimates align with observed rates of default for all groups. Consequently, we would expect use of an aware risk model to mitigate the observed gaps in group profits, through changes in both loan approvals and pricing. 
In Figure~\ref{si:approval-apr-changes}, we show that under this aware model, approval rates would decrease for Black borrowers ($-3$pp) and men ($-0.5$pp), and correspondingly increase for other racial and ethnic groups ($+1$pp) and women ($+1.5$pp). Average APRs would remain relatively unchanged for White, Hispanic, Asian, and women borrowers, but would increase by 0.8 points for Black borrowers and 0.3 points for men. We include these results for illustrative purposes only, and note that the use of a race- or gender-aware underwriting model would be legally impermissible under current fair lending law.

These results highlight a tension between the original policy goals of fair lending legislation and the realities under algorithmic pricing. 
When ECOA and FHA were written, discrimination concerns stemmed primarily from the individual discretion that was afforded to loan officers in making lending decisions and setting loan terms. 
In that context, prohibiting the use of protected characteristics in decision-making was a natural approach for promoting fair treatment of women and minority borrowers. 
Yet, in the context of algorithmic pricing, excluding protected characteristics from underwriting models may contribute to unfair lending outcomes if blind models are systematically miscalibrated across groups. 
Our analysis shows that using race- and gender-aware estimates of risk would eliminate observed group pricing disparities, relative to true creditworthiness. 
However, doing so would go directly against the legal constraints set forth in ECOA and FHA that prohibit the use of protected characteristics in decision-making. 
While other approaches for mitigating the miscalibration may exist, these solutions may be difficult to implement in practice. 
For example, one might try identifying additional predictive factors that, when included, eliminate the gap---but it is not clear what those factors may be or how feasible it would be to collect them.
The result is a tension in competing notions of fairness: race- and gender-blind decision making on the one hand, or similar pricing for similarly creditworthy individuals on the other.

\subsection{Strategic Shopping and Price Discrimination}
\label{section:shopping}

In addition to risk aversion and model miscalibration, we consider a third possible explanation for the gaps in profitability we see: strategic shopping.
In particular, it may be the case that men and Black borrowers are more sophisticated shoppers and only accept loan offers that are relatively favorable. 
To test this theory, we make use of data on the initial offers presented to applicants.
We then estimate counterfactual group IRRs in the absence of any strategic shopping---that is, assuming applicants accept any loan they were approved for. 

To estimate counterfactual loan profits in the absence of strategic shopping behavior, we simulated the scenario in which each loan applicant accepts the initial loan that they requested from the lender, if they are approved. In this scenario, loan approval implies loan origination. That is, applicants do not walk away from loans they have been approved for or request and accept a counteroffer for a different loan amount or APR.

To determine the counterfactual group IRRs of this no-shopping scenario, we trained a model to predict the IRR of individual loans using the applicant's race-and gender-aware risk score\footnote{We used aware risk scores to mitigate model miscalibration, as our goal was to estimate outcomes, not inform lending decisions.}, the loan amount, and APR. 
Because loan IRRs are only realized for individuals who both were approved and accepted their loan offer, we train our counterfactual IRR model on the set of funded borrowers---specifically, the subset of funded borrowers for whom the loan amount they requested equaled the loan amount that they ultimately received. We acknowledge that the population of funded borrowers may differ from the population of approved applicants, and that our estimated counterfactual IRRs for some applicants may involve some extrapolation.
Specifically, we fit a linear regression model of the following form: 
$$\text{IRR}_i = \alpha R_{A,i} + \gamma \text{LoanAmount}_i + \beta \text{APR}_i$$ 
where $\text{IRR}_i$ is the realized annualized IRR of the loan made to applicant $i$, $R_{A,i}$ is the aware risk score for applicant $i$ with coefficient $\alpha$, $\text{LoanAmount}_i$ is the amount of the loan made to applicant $i$ with coefficient $\gamma$, and $\text{APR}_i$ is the APR of the loan made to applicant $i$ with coefficient $\beta$. 

\begin{figure}
\centering
\includegraphics[width=0.6\linewidth]{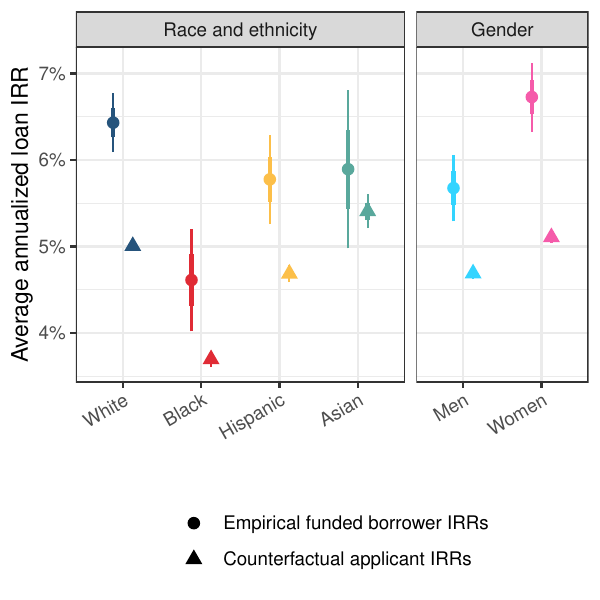}
\caption{\textbf{Counterfactual and realized internal rates of return (IRRs).} Counterfactual IRRs are estimated for loan applicants under the assumption that all approved applicants accept their first offer, removing effects of strategic borrower shopping and loan selection. Realized IRRs are calculated from the actual set of funded loans. Both sets of estimates are calculated by averaging IRRs of individual loans. Similar to the realized IRRs, the counterfactual IRRs of loans to men and Black borrowers are lower relative to other groups, suggesting that the profit gaps we observe are not solely attributable to potential differences in shopping behavior across groups.}
\label{fig:shopping}
\end{figure}

Then, using data on applicants approved for 3-year loans, we obtained counterfactual IRRs by applying our fitted model, assuming that each approved applicant accepted the loan they initially requested. Finally, we then used the BISG predictions to compute a weighted average of individual loan IRRs within each group, as shown by the triangular points in Figure \ref{fig:shopping}. In contrast, the IRR estimates shown in Figure \ref{fig:irr} were obtained by computing the IRR of a single loan portfolio aggregated at the group level, as opposed to averaging individual loan IRRs as we did in this counterfactual analysis. Therefore, to facilitate a more direct comparison, we computed the average of individual loan IRRs of funded borrowers (the same set of borrowers used to generate Figure \ref{fig:irr}), shown by the circular points in Figure \ref{fig:shopping}.

Figure \ref{fig:shopping} compares these estimated counterfactual IRRs (triangular points)---absent any shopping---to the real IRRs we observe for actually funded borrowers (circular points).\footnote{Note that the empirical IRR estimates differ slightly from those in Figure \ref{fig:irr} because the IRR estimates in Figure \ref{fig:irr} represent the IRR of each group's aggregate loan portfolio, whereas the estimates in Figure \ref{fig:shopping} represent a weighted average of individual loan IRRs within each group.} 
The counterfactual IRRs mirror the real IRRs, with Black borrowers generating significantly lower profits than other racial and ethnic groups, and men generating lower profits than women. 
We note that the counterfactual IRRs are estimated to be lower than the real IRRs because, if every applicant were to accept their loan offer, the lender would get a lower risk portfolio without losing the best borrowers to other lenders. In reality, lower risk borrowers often walk away or shop for better terms, leaving the lender with a riskier set of borrowers and higher overall profits associated with the added risk they are taking on. 
Ultimately, the similar patterns between the counterfactual and real group IRRs suggest that the profit gaps we observe are not explained by group differences in shopping behavior.

Finally, we note that the types of profit gaps we observe could, in theory, arise from price discrimination. For example, under a profit-maximizing strategy, a lender might seek to price loans higher for borrowers whom they perceive to have a higher willingness-to-pay. If loan demand elasticities correlate with protected group status, facially neutral price discrimination could consequently result in differential profits across groups. However, courts have specifically rejected ``market forces'' as a legitimate business justification for disparities in lending outcomes \cite{MillerCountrywide2008} and have maintained that business necessity under ECOA means a manifest relationship to creditworthiness\cite{ABAutoService1997, LewisACB1998}. Therefore, profit disparities across protected groups arising from price discrimination would likely constitute a legally impermissible form of disparate impact. In this sense, price discrimination is not a source of confounding in a profit-based test; rather, it is exactly the type of conduct that this approach is designed to surface.

\section{Discussion}

Auditing approximately 80,000 loans originated by a major U.S. fintech platform, we found that loans to Black borrowers and men yielded lower profits than loans made to White, Hispanic, Asian and women borrowers. 
Our evidence suggests that this result is likely driven by miscalibration in the risk model used by the platform, which underestimates the default risk of Black borrowers and overestimates the default risk of women, leading the platform to inaccurately price loans for these groups. 
In turn, conditional on true riskiness, Black borrowers in our sample receive more favorable interest rates relative to other racial and ethnic groups, and women in our sample receive less favorable interest rates relative to men, on average. 

Under fair lending statutes like the Equal Credit Opportunity Act of 1974, such risk-independent disparities could constitute a form of impermissible disparate impact. 
As we show, using a race- and gender-aware underwriting model could mitigate the gap, although doing so would constitute a form of legally impermissible race- and gender-based decision making, or disparate treatment \cite{pope2011implementing}. This tension echoes recent legal scholarship calling for regulatory frameworks better tailored to the oversight of algorithmic pricing systems used in credit markets \cite{Gillis2022, gillis2024price}.

Employing an IRR-based test of discrimination is a practical and straightforward regulatory approach for monitoring lenders that employ algorithmic underwriting, especially in settings where researchers may not have access to the full set of risk inputs used by the lender. This approach aligns with recent work in the algorithmic fairness literature advocating for a shift towards outcome-based approaches to evaluating the fairness of algorithms \cite{Gillis2022, ChohlasWood2023}. 
It also addresses the well-known limitations of adverse impact ratios, the 
problem of omitted-variable bias in benchmark tests of discrimination~\cite{ayres2002outcome}, and the problem of inframarginality in tests based on default rates~\cite{simoiu2017problem}. 
Its simplicity---requiring only applicant demographics and loan cash flow vectors---also means that it can be easily generalized to other credit markets and serve as a useful first indicator of whether additional regulatory scrutiny is warranted. 

Our analysis is subject to several important limitations. 
First, a profit-based test is unable to detect all forms of potential discrimination. For example, a lender might hypothetically refuse to lend to a random subset of minority applicants, but then offer non-discriminatory loan terms to those who are ultimately offered loans. Such discriminatory behavior would not be detected by a comparison of profits across groups. Our audit is therefore limited to assessing discrimination in loan pricing, as opposed to approval decisions. Second, current legislation prohibits lenders from collecting protected characteristics of non-mortgage applicants, and so our analysis relies on the use of imputed demographics~\cite{USCongress1974}. 
Recent work has suggested that using such estimates in studies of outcome disparities may significantly underestimate the magnitude of racial disparities, in which case, the disparities we report herein may be greater than we estimate \cite{chernenko2023limits}.
It is worth noting, however, that there is precedent for using BISG estimates by regulators, such as the Consumer Financial Protection Bureau and the Department of Justice \cite{cfpb2014_bisg, Akinwumi2021, CFPB_DOJ_Ally_Consent_2013}.

Our analysis focused on personal loans originated by a single U.S. fintech lender, but the findings likely generalize to other lenders and credit markets. For example, recent work uncovered similar miscalibration in consumer credit scores for Black borrowers \cite{bakker2025credit} and other work observed underestimation of delinquency risk in business credit scores for minority business owners \cite{robb2018testing}. Other recent work scrutinizing the calibration of consumer credit scores across gender in the subprime borrowing context similarly found that the default risk for women is overestimated \cite{LIU2025101081}. This pattern suggests that, in credit markets where these credit scores are used,  other audits could reveal analogous disparities to those we uncover in our analysis. Nevertheless, the magnitude and direction of disparities may vary across borrower populations and credit products, and  replication of our analyses in different credit contexts would be a valuable next step in assessing the prevalence of the patterns we observe.

Our work illustrates that even facially neutral algorithms can create consequential racial and gender disparities, and
underscores the importance of predictive accuracy in the pursuit of equitable lending~\cite{blattner2021costly}. 
We hope that our analysis assists researchers and policymakers in the study, regulation, and remediation of lending discrimination.

\section*{Acknowledgments}
M.~Coots was supported by a James M. and Cathleen D. Stone PhD Scholar fellowship from the Stone Program in Wealth Distribution, Inequality, and Social Policy at Harvard University, and by a Social Equity and Health Equity stipend from the Malcolm Wiener Center for Social Policy at Harvard Kennedy School. The authors thank
Desmond Ang, 
Susan Athey, 
Matthew Baum, 
Deirdre Bloome, 
Lucas Chancel, 
Adam Chilton,
Johann Gaebler, 
Talia Gillis, 
Jacob Jameson,
Taeku Lee, 
Daniel Schneider, 
Maya Sen, 
Max Spohn, and
Richard Zeckhauser
for helpful conversations and feedback.

\clearpage

\bibliography{refs}

\clearpage

\part*{Appendix}

\appendix
\renewcommand{\thefigure}{A\arabic{figure}}
\setcounter{figure}{0}

\section*{BISG Robustness Check}
\label{si:bisg-robustness}

In line with recommendations from the Consumer Financial Protection Bureau, our main results were obtained by using the BISG probability estimates as weights throughout our analyses \cite{cfpb2014_bisg}. However, more recent work has recommended using inferred labels on race and gender status by taking the argmax of BISG probability estimates  \cite{zhang2018assessing}. As a robustness check on our results, we regenerated our main results using this approach. Concretely, $R \in \{\text{Asian}, \text{Black}, \text{Hispanic}, \text{White}\}$ was assigned for each individual as 

$$R = \operatorname*{arg\,max}_R \, \hat{p}_R$$

\noindent and $G \in \{\text{Woman}, \text{Man}\}$ was assigned for each individual as 

$$G = \operatorname*{arg\,max}_G\, \hat{p}_G$$ where $\hat{p}_R$ and $\hat{p}_G$ denote the estimated probability distributions of membership in race and gender groups, respectively.

\begin{figure}[th]
\centering
\includegraphics[width=8cm]{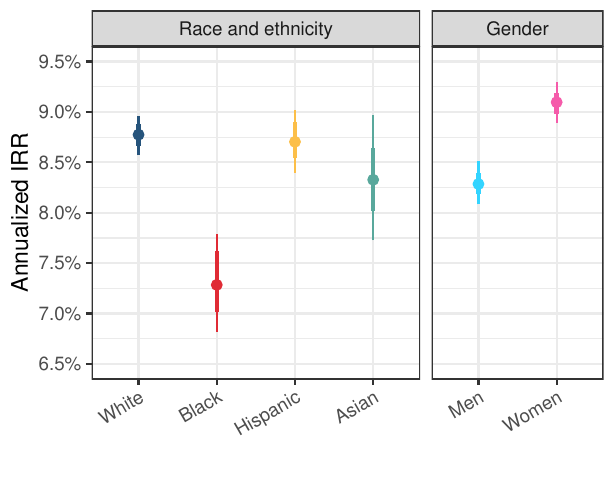}
\caption{\textbf{Robustness check of Figure \ref{fig:irr} using the argmax of BISG estimates  to obtain race and gender labels, as opposed to using the BISG estimates as weights.} The group IRR estimates produced under the argmax method are qualitatively similar to those shown in Figure \ref{fig:irr} in the main text.}
\label{si:irr-bisg-max}
\end{figure}

\begin{figure}[th]
\centering
\includegraphics[width=10cm]{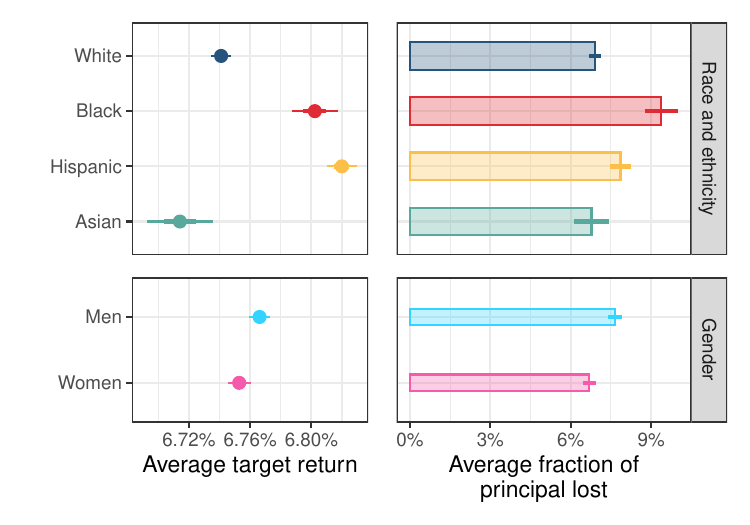}
\caption{\textbf{Robustness check of Figure \ref{fig:target_return_principal_lost} using the argmax of BISG estimates to obtain race and gender labels, as opposed to using the BISG estimates as weights.} Both sets of results produced under the argmax method are qualitatively similar to those shown in Figure \ref{fig:target_return_principal_lost} in the main text.}
\label{si:target-return-principal-lost-bisg-max}
\end{figure}

\begin{figure}[th]
\centering
\includegraphics[width=10cm]{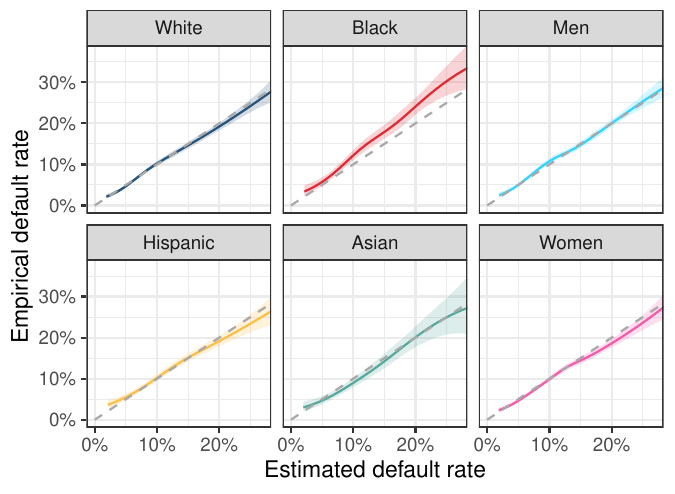}
\caption{\textbf{Robustness check of Figure \ref{fig:blind-calib} using the argmax of BISG estimates to obtain race and gender labels, as opposed to using the BISG estimates as weights.} The group calibration results produced under the argmax method are qualitatively similar to those shown in Figure \ref{fig:blind-calib} in the main text.}
\label{si:blind-calib-bisg-max}
\end{figure}

\begin{figure}[th]
\centering
\includegraphics[width=8cm]{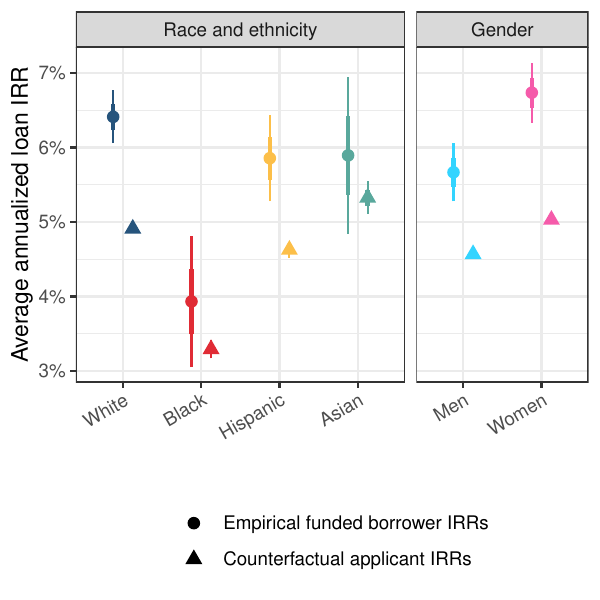}
\caption{\textbf{Robustness check of Figure \ref{fig:shopping} using the argmax of BISG estimates to obtain race and gender labels, as opposed to using the BISG estimates as weights.} These results produced under the argmax method are qualitatively similar to those shown in Figure \ref{fig:shopping} in the main text.}
\label{si:shopping-bisg-max}
\end{figure}

In Figures \ref{si:irr-bisg-max}, \ref{si:target-return-principal-lost-bisg-max}, \ref{si:blind-calib-bisg-max}, and \ref{si:shopping-bisg-max} we offer versions of Figures \ref{fig:irr}, \ref{fig:target_return_principal_lost},  \ref{fig:blind-calib}, and \ref{fig:shopping} generated using the above approach. 


\begin{figure}[th]
\centering
\includegraphics[width=8cm]{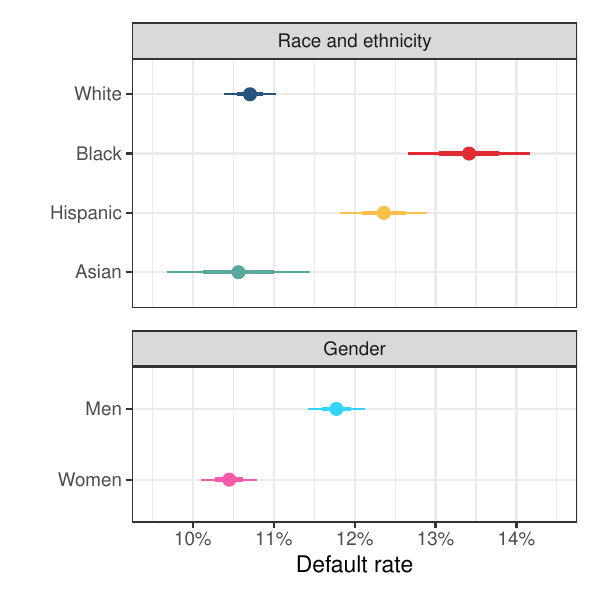}
\caption{\textbf{Rates of loan default computed across race and gender groups.} Black and male borrowers tend to be higher risk groups, as exhibited by their relatively higher rates of default. We show the 68\% confidence interval (thick bar) and the 95\% confidence interval (thin bar) of the default rate estimates. These estimates were computed using the set of funded borrowers.}
\label{fig:default-rate}
\end{figure}

\begin{figure}
\centering
\includegraphics[width=13cm]{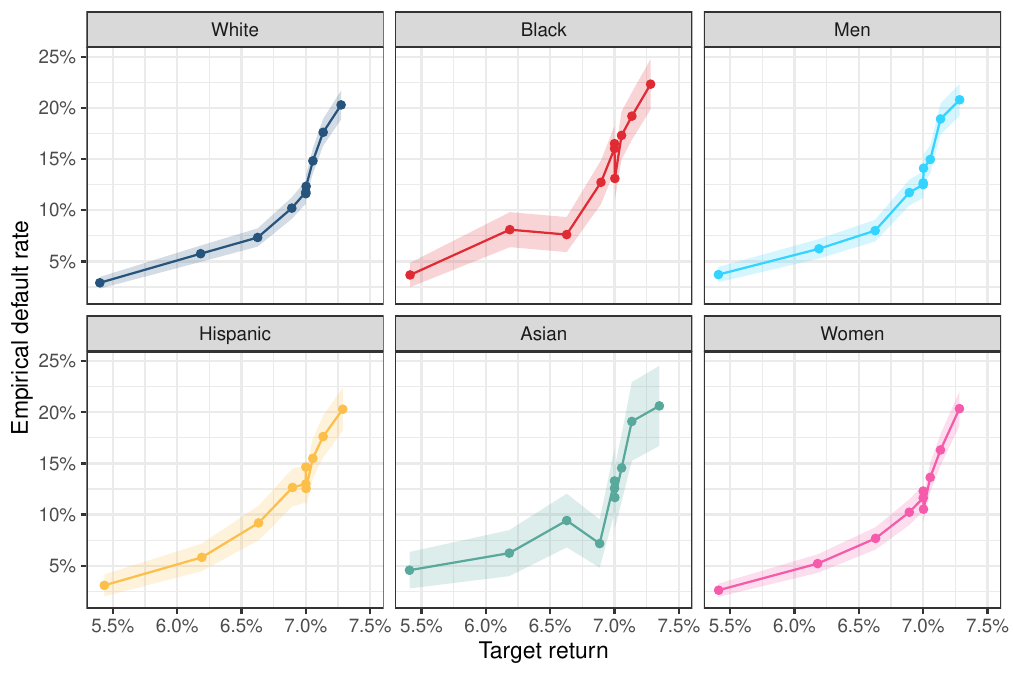}
\caption{\textbf{An empirical assessment of the relationship between target returns and default risk.} Curves were generated by computing the average target return and default rate within each group deciles of target return. The shaded area shows the 95\% confidence interval of our estimates. Across groups, target default risk exhibits a strong, positive correlation with target returns. These results were computed using the set of funded borrowers.
}
\label{si:default-target-returns}
\end{figure}

\begin{figure}
\centering
\includegraphics[width=8cm]{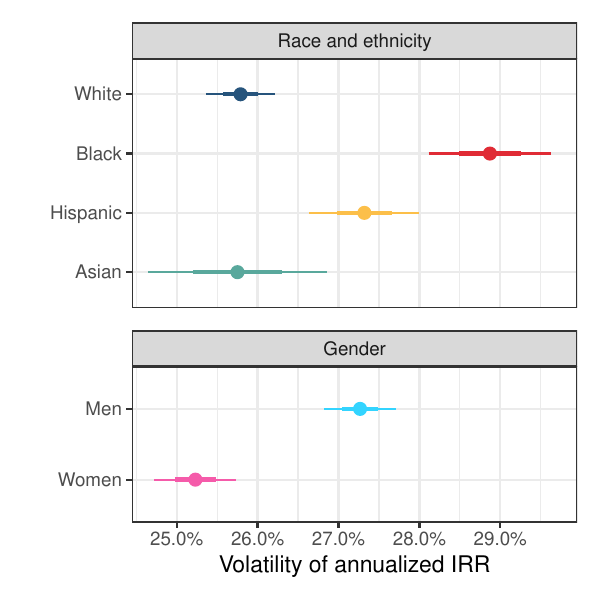}
\caption{\textbf{Volatility of loan IRRs computed across race and gender groups.} Black and male borrowers tend to be higher risk groups, as exhibited by their higher volatility of returns. We show the 68\% confidence interval (thick bar) and the 95\% confidence interval (thin bar) of the volatility estimates. These estimates were computed using the set of funded borrowers.}
\label{si:volatility}
\end{figure}

\begin{figure}
\centering
\includegraphics[width=10cm]{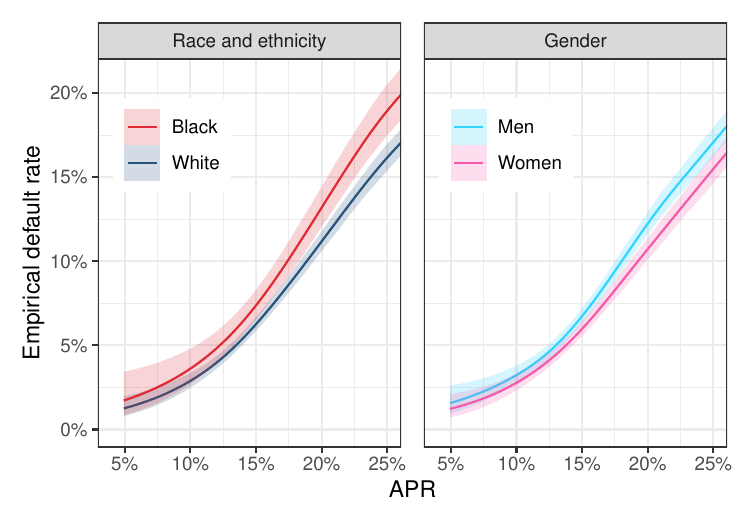}
\caption{\textbf{An empirical assessment of the calibration of the lender's internal risk score.} (Left) Given a fixed APR, Black borrowers default at higher rates than White borrowers. (Right) Similarly, given a fixed APR, women default at slightly lower rates than men. For each curve, we show the 95\% confidence interval. Curves were smoothed using a generalized additive model using a logit link function. These results suggest that the lender's internal risk score exhibits miscalibration similar to that which we observe in our own blind risk model, resulting in relatively favorable APRs for Black borrowers relative to White borrowers, and slightly unfavorable APRs for women relative to men.}
\label{si:interest-rate-fig}
\end{figure}

\begin{figure}
\centering
\includegraphics[width=10cm]{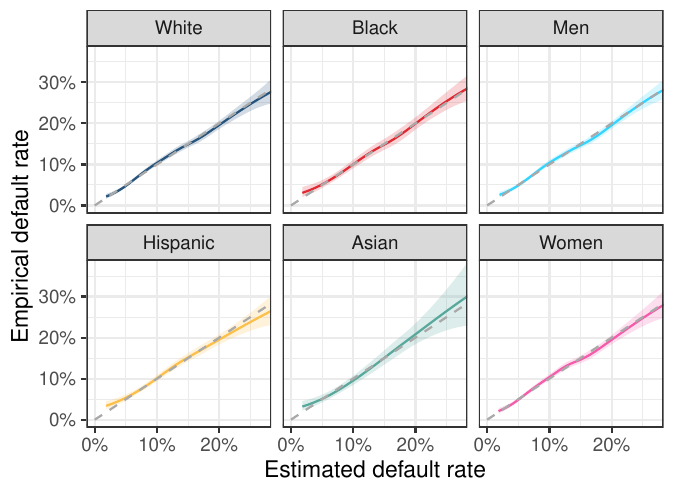}
\caption{\textbf{The calibration of reconstructed race- and gender-aware risk scores.} The line $y=x$ denotes the line of perfect calibration, shown by a dashed gray line. (Left and Middle) Aware risk scores are calibrated for all racial and ethnic subgroups. (Right) Similarly, aware risk scores are calibrated for both men and women. For each curve, we show the 95\% confidence interval. Curves were smoothed using a generalized additive model using a logit link function. These results suggest that an aware risk score would correct the inaccuracies in loan pricing stemming from the use of a blind risk score.}
\label{si:aware-calib}
\end{figure}

\begin{figure}
\centering
\includegraphics[width=11.4cm]{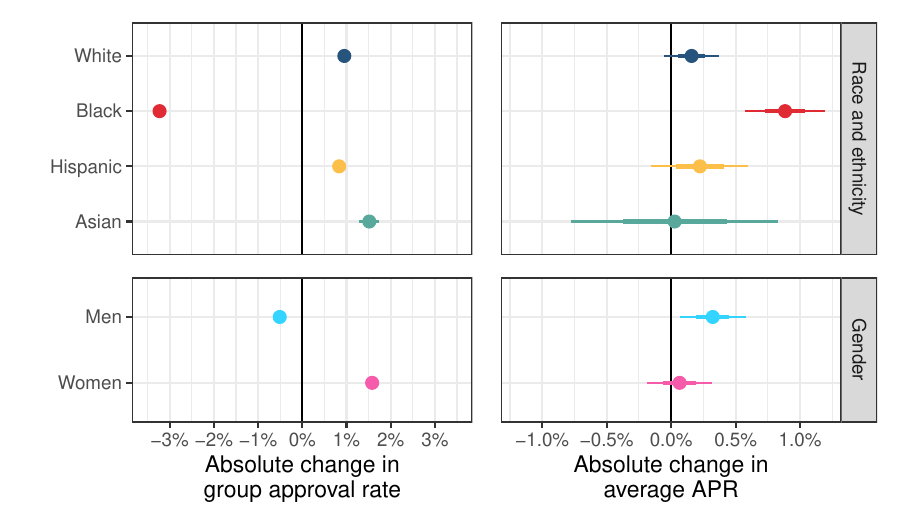}
\caption{\textbf{The predicted consequences of switching to a race- and gender-aware risk model for loan underwriting.} (Left) When disaggregated by race and ethnicity, loan approval rates would be expected to decrease for Black borrowers by more than 3pp, and increase by approximately 1pp for White, Hispanic, and Asian borrowers. By gender, approval rates would be expected to decrease for men by about 0.5pp and increase for women by approximately 1.5pp. (Right) Across racial and ethnic groups, loan APRs would increase for Black borrowers, but remain relatively unchanged for White, Hispanic, and Asian borrowers. By gender, APRs would increase for men, but remain constant for women. We show the 68\% confidence interval (thick bar) and the 95\% confidence interval (thin bar) for each estimate.}
\label{si:approval-apr-changes}
\end{figure}

\begin{figure}[th]
\centering
\includegraphics[width=8cm]{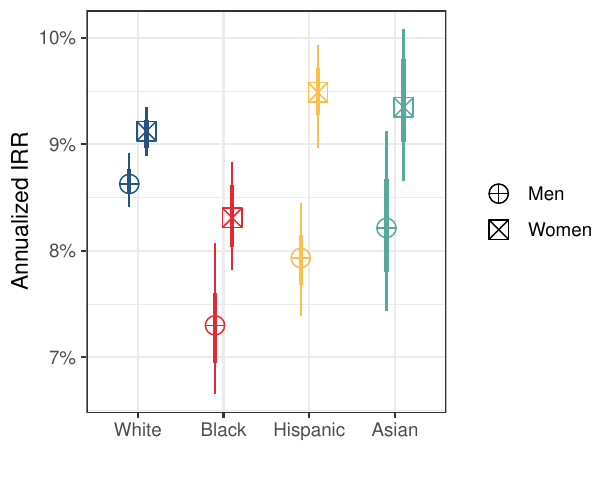}
\caption{\textbf{Results of computing the annualized IRR of cashflows aggregated across intersectional race-gender groups.} We show the 68\% confidence interval (thick bar) and the 95\% confidence interval (thin bar) of the IRR estimates.}
\label{si:intersection}
\end{figure}


\begin{table}
	\centering
    \small
	\caption{\textbf{Summary statistics for the population of funded borrowers.}
		All borrowers in this sample took out 3-year loans that were originated in 2019. Estimates for each subgroup were obtained using estimated weights for race and gender provided by the lender. 95\% confidence intervals are shown below each estimate. Individuals with race and/or gender weights that were all zero were labeled as unknown. For these individuals, the lender was unable to infer group weights based on the borrower's name. \\}
	\label{tab:funded_borrowers}

\begin{tabular}{llllll}
\toprule
Group & N (eff.) & Pct. & Default rate & Avg. loan amount (\$) & Avg. APR\\
\midrule
White & 38,070 & 48.0\% & \makecell{10.7\% \\ (10.4, 11.0)} & \makecell{9,402 \\ (9,326, 9,477)} & \makecell{19.4\% \\ (19.4, 19.5)}\\
Black & 8,234 & 10.4\% & \makecell{13.4\% \\ (12.9, 14.0)} & \makecell{8,668 \\ (8,542, 8,793)} & \makecell{20.2\% \\ (20.1, 20.3)}\\
Hispanic & 14,982 & 18.9\% & \makecell{12.4\% \\ (11.9, 12.8)} & \makecell{8,459 \\ (8,357, 8,562)} & \makecell{20.3\% \\ (20.2, 20.4)}\\
Asian & 4,795 & 6.1\% & \makecell{10.6\% \\ (9.8, 11.3)} & \makecell{11,334 \\ (11,106, 11,561)} & \makecell{19.1\% \\ (18.9, 19.2)}\\
American Indian & 356 & 0.4\% & \makecell{12.1\% \\ (10.7, 13.5)} & \makecell{8,793 \\ (8,456, 9,130)} & \makecell{20.0\% \\ (19.8, 20.3)}\\
\addlinespace
Multirace & 1,241 & 1.6\% & \makecell{11.3\% \\ (10.9, 11.7)} & \makecell{9,374 \\ (9,264, 9,484)} & \makecell{19.7\% \\ (19.6, 19.8)}\\
Unknown Race/Ethnicity & 11,572 & 14.6\% & \makecell{10.9\% \\ (10.4, 11.5)} & \makecell{10,335 \\ (10,177, 10,493)} & \makecell{19.3\% \\ (19.2, 19.4)}\\
\midrule
Women & 31,383 & 39.6\% & \makecell{10.5\% \\ (10.1, 10.8)} & \makecell{9,418 \\ (9,330, 9,506)} & \makecell{19.6\% \\ (19.5, 19.6)}\\
Men & 34,375 & 43.4\% & \makecell{11.8\% \\ (11.4, 12.1)} & \makecell{9,077 \\ (8,993, 9,160)} & \makecell{19.7\% \\ (19.6, 19.8)}\\
Unknown Gender & 13,493 & 17.0\% & \makecell{12.3\% \\ (11.7, 12.8)} & \makecell{10,165 \\ (10,023, 10,307)} & \makecell{19.7\% \\ (19.6, 19.8)}\\
\addlinespace
\midrule
All Funded Borrowers & 79,251 & 100.0\% & \makecell{11.3\% \\ (11.1, 11.6)} & \makecell{9,397 \\ (9,341, 9,454)} & \makecell{19.6\% \\ (19.6, 19.7)}\\
\bottomrule
\end{tabular}
\end{table}

\begin{table} 
	\centering
	\caption{\textbf{Summary statistics for the population of loan applicants.}
		This sample includes individuals who applied for 3-year loans between 2021 and 2023, and includes a mix of both approved and denied applicants, as well as approved applicants who did not ultimately accept a loan offer from the lender. Estimates for each subgroup were obtained using estimated weights for race and gender provided by the lender. 95\% confidence intervals are shown below each estimate. Individuals with race and/or gender weights that were all zero were labeled as unknown. For these individuals, the lender was unable to infer group weights based on the borrower's name. The offer acceptance rate refers to the proportion of approved applicants that accepted the lender's loan offer. Average accepted loan amount and APR were computed using the set of loans both approved by the lender and accepted by the applicant.\\}
	\label{tab:applicants} 

\resizebox{\textwidth}{!}{
\begin{tabular}{llllll}
\toprule
Group & N (eff.) & Pct. & Offer acceptance rate & Avg. accepted loan amount (\$) & Avg. accepted APR\\
\midrule
White & 63,762 & 50.9\% & \makecell{13.4\% \\ (13.0, 13.7)} & \makecell{8,655 \\ (8,585, 8,726)} & \makecell{25.2\% \\ (25.1, 25.2)}\\
Black & 22,819 & 18.2\% & \makecell{16.4\% \\ (15.7, 17.0)} & \makecell{7,470 \\ (7,375, 7,566)} & \makecell{26.6\% \\ (26.6, 26.7)}\\
Hispanic & 19,815 & 15.8\% & \makecell{18.8\% \\ (18.0, 19.6)} & \makecell{7,512 \\ (7,398, 7,626)} & \makecell{25.9\% \\ (25.8, 26.0)}\\
Asian & 3,439 & 2.7\% & \makecell{14.5\% \\ (13.1, 15.8)} & \makecell{11,399 \\ (11,071, 11,728)} & \makecell{22.9\% \\ (22.7, 23.1)}\\
American Indian & 788 & 0.6\% & \makecell{14.7\% \\ (12.7, 16.8)} & \makecell{7,477 \\ (7,155, 7,798)} & \makecell{26.3\% \\ (26.1, 26.6)}\\
\addlinespace
Multirace & 1,978 & 1.6\% & \makecell{14.6\% \\ (14.1, 15.2)} & \makecell{8,480 \\ (8,380, 8,580)} & \makecell{25.5\% \\ (25.4, 25.6)}\\
Unknown Race/Ethnicity & 12,744 & 10.2\% & \makecell{13.6\% \\ (12.7, 14.5)} & \makecell{9,182 \\ (9,001, 9,363)} & \makecell{24.8\% \\ (24.7, 24.9)}\\
\midrule
Women & 54,130 & 43.2\% & \makecell{15.4\% \\ (14.9, 15.9)} & \makecell{7,861 \\ (7,784, 7,938)} & \makecell{25.6\% \\ (25.5, 25.7)}\\
Men & 51,057 & 40.7\% & \makecell{13.8\% \\ (13.4, 14.3)} & \makecell{9,067 \\ (8,977, 9,157)} & \makecell{25.2\% \\ (25.1, 25.2)}\\
Unknown Gender & 20,158 & 16.1\% & \makecell{15.6\% \\ (14.7, 16.4)} & \makecell{8,018 \\ (7,884, 8,152)} & \makecell{25.7\% \\ (25.6, 25.8)}\\
\addlinespace
\midrule
All Loan Applicants & 125,345 & 100.0\% & \makecell{14.7\% \\ (14.4, 15.1)} & \makecell{8,378 \\ (8,323, 8,432)} & \makecell{25.4\% \\ (25.4, 25.5)}\\
\bottomrule
\end{tabular}
}
\end{table}

\end{document}